\documentclass[usegraphicx,usenatbib,useAMS]{mn2e}
\usepackage{times}
\usepackage{amssymb}
\def\gs{\mathrel{\raise1.16pt\hbox{$>$}\kern-7.0pt %
\lower3.06pt\hbox{{$\scriptstyle \sim$}}}}         %
\def\ls{\mathrel{\raise1.16pt\hbox{$<$}\kern-7.0pt %
\lower3.06pt\hbox{{$\scriptstyle \sim$}}}}         %

\begin{document}

\title[]{The star formation histories of galaxies in the Sloan Digital Sky Survey}
\author[Panter et al.]
{Benjamin Panter$^{1,2}$, Raul Jimenez$^3$, Alan F. Heavens$^2$ and Stephane Charlot$^{1,4}$ \\
$^1$Max-Planck-Institut fur Astrophysik, Karl-Schwarzschild Str. 1, D-85748, Garching bei Munchen, Germany\\
$^2$SUPA \thanks{Scottish Universities Physics Alliance}, Institute
for Astronomy, University of Edinburgh, Royal
Observatory, Blackford Hill, Edinburgh EH9 3HJ, UK\\
$^3$Dept. of Physics and Astronomy, University of Pennsylvania,
Philadelphia,
PA 19104, USA\\
$^4$Institut d'Astrophysique de Paris, UMR 7095, 98 bis Boulevard
Arago, F-75014 Paris, France}

\maketitle

\begin{abstract}
We present the results of a MOPED analysis of $\sim 3 \times 10^5$
galaxy spectra from the Sloan Digital Sky Survey Data Release Three
(SDSS DR3), with a number of improvements in data, modelling and
analysis compared with our previous analysis of DR1.  The
improvements include: modelling the galaxies with theoretical models
at a higher spectral resolution of 3\AA; better calibrated data; an
extended list of excluded emission lines, and a wider range of dust
models. We present new estimates of the cosmic star formation rate,
the evolution of stellar mass density and the stellar mass function
from the fossil record. In contrast to our earlier work the results
show no conclusive peak in the star formation rate out to a redshift
around 2 but continue to show conclusive evidence for `downsizing'
in the SDSS fossil record. The star formation history is now in good
agreement with more traditional instantaneous measures. The galaxy
stellar mass function is determined over five decades of mass, and
an updated estimate of the current stellar mass density is presented.
We also investigate the systematic effects of changes in the stellar
population modelling, the spectral resolution, dust modelling, sky
lines, spectral resolution and the change of data set. We find that
the main changes in the results are due to the improvements in the
calibration of the SDSS data, changes in the initial mass function
and the theoretical models used.
\end{abstract}

\begin{keywords}
  galaxies: fundamental parameters, galaxies: statistics, galaxies:
  stellar content
\end{keywords}

\section{Introduction}

The quality of spectra of the observed light of unresolved stellar
populations has reached enough accuracy that it is possible to make
detailed studies of the physical properties of the stellar
populations in these galaxies. An excellent example of this new
generation of data-sets is given by the Sloan Digital Sky Survey
\citep{Gunn+98,York+00,Strauss+02} at low redshift, not only by the
size of the spectroscopic sample (about $10^6$ spectra) but by the
quality and wavelength coverage of the spectra. At higher redshift
the DEEP2 survey \citep{Davis+03} is providing a similar database,
albeit with a  smaller wavelength coverage. Future spectroscopic
surveys (e.g. WFMOS) will yield larger samples at even deeper
redshifts. Given the quality of the spectra, it is interesting to
ask the question of whether reliable information about the stellar
population of these galaxies can be inferred from the spectra.

Indeed, several attempts have been made previously to study in
detail the physical properties of the SDSS galaxies either by using
selected features in the spectra \citep{Kauffmann+04,Brinchmann04,
Tremonti+04} or using the full spectrum \citep{ PHJ03,HPJD04,PHJ04,
Cid05,Mathis06,Ocvirk06}. These studies have led to interesting
conclusions about the physical properties of these galaxies. In
particular analysis of the SDSS sample \citep{HPJD04} and other
local galaxies \citep{Thomas05} show very clear evidence for
`downsizing' - the process by which star formation at low redshift
takes place predominantly in low-mass galaxies, whereas more massive
galaxies have the bulk of their star-formation activity at high
redshift. also observed using other methods. In addition one can
also determine the global star formation history in the Universe
from this fossil record.  Our previous study found broad agreement
with observations of contemporary star formation at large lookback
times, but suggested that there was a peak in star formation
activity at $z<1$. Agreement between the fossil record and
contemporary star formation indicators is expected if the
Cosmological Principle holds; this expected agreement offers an
opportunity in principle to test the assumptions in the modelling of
both the fossil record and the instantaneous star formation rates.
Of most importance is that both are dependent on the stellar initial
mass function, but in different ways, with the instantaneous rates
being determined very much by the upper end of the IMF, while the
fossil record is determined by a wider range of stars.  This offers
an opportunity in principle to test the evolution of the IMF as a
function of cosmic time.

In our previous work, we have presented results from around $10^5$
galaxies from the SDSS DR1.  This paper enhances the previous
results through the introduction of a number of improvements to data
and analysis methods, and through an investigation of systematic
effects and sensitivity to assumptions.  On the data side, we have
analysed $\sim 3 \times 10^5$ galaxies from the SDSS DR3 sample. The
extra size of this sample is not particularly important, but the
calibration of the data has been improved since DR1.  On the methods
side, we no longer rebin the data to 20\AA\ resolution, but compare
the data with a wider range of theoretical models, including the
Bruzual \& Charlot models \citep{BC03} at 3\AA\ resolution.  We
consider two stellar initial mass functions (a Salpeter and a
Chabrier IMF), and a wider range of dust models. We also make some
improvements to the treatment of emission lines, by removing
additional weaker lines, since interstellar emission lines are not
included in the stellar modelling.  Finally, we have also explored
the effect of removal of sky lines, using a PCA-based method
\citep{Wild05}. These studies give us a reasonable estimate of the
sensitivity of the results to the assumptions.  As is expected from
a sample of this size, changes in the assumptions give rise to much
larger variations in the results than the statistical errors. The
uncertainties in star formation rates from the stellar models used
are typically about a factor of two, which makes it difficult to
distinguish between different IMFs, although it may be possible to
constrain extreme IMF variations.

The layout of the paper is as follows.  In Section 2 we outline the
enhancements to the method, and the improvements to the data, and
describe the assumptions which are made for analysis of the DR3
sample.  In Section 3 we present the new results on star formation
history, downsizing and galaxy stellar mass function from the DR3
dataset. We also introduce a method to assess the particular areas
where current models are lacking. In Section 4 we show the
sensitivity of our results to changes in the assumptions, through
analysis of a subset of the data, and in Section 5 we draw
conclusions.

Throughout this work we assume a concordance cosmology with
$\Omega_v=0.73$, $\Omega_m=0.27$, $H_0=71$
kms$^{-1}$Mpc$^{-1}$\citep{SpergelWMAP03}.

\section{SDSS DR3 analysis}

In this Section we describe briefly the SDSS DR3 dataset, and
outline the assumptions in the method, highlighting changes made
since the analysis of SDSS DR1 \citep{HPJD04}.

\subsection{SDSS DR3 data}

The spectrophotometric pipeline used by the SDSS has evolved from
the Early Data Release (EDR) to the DR3 sample used here. According
to the papers describing each release \citep{EDR02, SDSS-DR1,
SDSS-DR2, SDSS-DR3}, the pipeline was changed between the EDR, DR1
and DR2 subsets, but remained static between DR2 and DR3. Some DR1
spectra had a slight systematic offset for wavelengths smaller than
$4000 $\AA. \citet{SDSS-DR3} claim that this has been corrected in
the DR3. There are no published plans for further improvements. The
DR1 data has been re-reduced with the new pipeline and this work
considers the set of galaxies contained in the SDSS Main Galaxy
Sample (MGS) of DR1-3 reduced with the DR2 version of the
spectrophotometric pipeline.

We apply further cuts to this main galaxy sample based on those of
\citet{Shen+03}. Our sample is determined by r band apparent
magnitude limits of $15.0 \le m_r \le 17.77$. The magnitude limits
are set by the SDSS target selection criteria, as discussed in
\citet{SDSS-DR3}. The target criteria for surface brightness was
$\mu_r<24.5$, although for $\mu_r>23.0$ galaxies are included only
in certain atmospheric conditions. In order to remove any bias and
simplify our $V_{\rm max}$ criteria we have cut our sample at
$\mu_r<23.0$. At low redshifts a small number of Sloan galaxies are
subject to shredding - where a nearby large galaxy is split by the
target selection algorithm into several smaller sources. To
eliminate this effect, for our star formation analysis we use a
range of $0.01<z<0.25$. This also removes the problems of non
Hubble-flow peculiar velocities giving erroneous distances based on
redshift, which can have a significant effect on recovered stellar
mass. For samples involving a very large redshift range there is a
concern that after $V_{\rm max}$ weighting an individual galaxy at
low redshift can dominate higher redshift galaxy signals. For our
criteria we have tested this, and no $V_{\rm max}$ weighted galaxy
contributes more than a tenth of a percent to the final mass total
in each bin. The total number of galaxies in the DR3 Main Galaxy
Catalogue is $312415$, while the number that satisfy our cuts is
$299571$. In order to estimate the completeness of the survey we
have used the ratio of target galaxies to those which have observed
redshifts (P. Nordberg, Priv. Comm.). This does not allow for
galaxies which are too close for the targetting algorithm, and we
estimate this fraction at a $6\%$ level from the discussion in
\citet{Strauss+02}. As a result of both these cuts, our effective
sky coverage is $2947$ square degrees.

We also remove from our analysis a larger set of wavelengths which
may be affected by emission lines than in our DR1 analysis. These
lines are not modelled by the stellar population codes, which only
consider the continuum and absorption features. The excluded
restframe wavelength ranges due to emission or emission line filling
of features are listed in table \ref{tab:emission}. We also discount
signal with wavelength above 7800\AA\ in order to reduce the risk of
skyline contamination. Typically the rest frame wavelength range
used for analysis is 3450\AA\ - 7800\AA.

\begin{table}
 \centering
 \begin{minipage}{\columnwidth}
  \caption{Regions masked from MOPED fitting.}
\begin{tabular}{|c|c|c|c|c|c|}
  \hline
Wavelength Range (\AA)& Reason\\
  \hline
3711-3741 & [OII]\\
4087-4117 & [SII], H$\delta$\\
4325-4355 & H$\gamma$\\
4846-4876 & H$\beta$\\
4992-5022 & [OIII]\\
4944-4974 & [OIII]\\
5870-5900 & Na\\
6535-6565 & [NII]\\
6548-6578 & H$\alpha$\\
6569-6599 & [NII]\\
6702-6732 & [SII]\\
6716-6746 & [SII]\\
  \hline
  \label{tab:emission}
\end{tabular}
\end{minipage}
\end{table}
\subsection{Modelling assumptions}

\subsubsection{Stellar population modelling}

Both our EDR \citep{PHJ03} and DR1 \citep{HPJD04,PHJ04} studies used
the 20 \AA\ resolution models described in \citet{JMDPP04}. The
field of stellar population modelling has moved on in the meantime,
and models at 3 \AA\ resolution and better are available from
various authors. We have used those of \citet{BC03} as the basis for
this study.

%Although this does reduce the signal to noise of corresponding
%rebinned galaxy spectra, the increase in information contained in the
%models more than compensates for this. If the Sloan spectra are
%considered to have a signal to noise of XX, then the 3 \AA spectra
%have a S/N of XX compared to the 20 \AA rebinned spectra with S/N of
%XX.

Although with 20\AA\ models the effect of velocity dispersion can be
ignored, there is possibility of significant changes in the input
spectrum when working at 3\AA. After extensive testing we found that
in fact there is very little effect, if any, on the recovered
stellar populations and metallicities for a wide range of galaxies.
We chose to apply a uniform velocity dispersion of 170 kms$^{-1}$ to
the 3\AA\ models, reflecting a typical value for the Main Galaxy
Sample.

\subsubsection{Initial Mass Functions}

Both the instantaneous and fossil approach to star formation
determination require assumptions about the Initial Mass Function
(IMF).  The two methods probe different mass regions and deduce the
complete stellar populations by assuming an IMF. For instantaneous
measures such as H-$\alpha$ or OII emission the presence of lower
mass stars is estimated by working down the mass function from the
massive stars which cause the majority of the emission. For
high-redshift star formation estimated using the fossil record
technique, the contribution to the spectrum from older, less massive
stars is used to determine early star formation, requiring
extrapolation up the mass function.

In recent years, the choice of IMF has been the subject of much
debate - in particular whether a universal IMF can be assumed, both
in space and time. Several candidates have been proposed for such a
IMF, however single stellar population (SSP) models only include a
few. Although disfavoured by observations, the Salpeter IMF
\citep{Salpeter} has been used as a reference due to its simplicity;
it is a power law. The most recent modification is the Chabrier IMF
\citep{Chabrier} which seems to be very successful at reproducing
current observations in our galaxy. For the main analysis, we use
the Chabrier IMF.

\subsubsection{Dust}

Our previous MOPED studies used a single foreground dust screen. In
this parameterisation the strength of the extinction may be
characterised by $E(B-V)$, and the wavelength dependence of the
extinction is determined by the choice of extinction model, which
may be empirical or modelled. We used an LMC extinction law for the
main analysis, but later we will explore the difference in recovered
SFH using the \citet{Calzetti97} starburst model and the LMC and SMC
curves given in \citet{Gordon03}.

We have also computed results using the two-dust parameter model of
\citet{CF00}. This is a more physically motivated model of the
absorption of starlight by dust, which accounts for the different
attenuation affecting young ($<10$ Myr) and old stars in galaxies,
as characterized by the typical absorption optical depths of dust in
giant molecular clouds and in the diffuse ISM. Unfortunately, our
investigations have shown that the absorption in giant molecular
clouds cannot be well constrained from the optical continuum
emission alone. This dust component would be more tightly
constrained by the ultraviolet and infrared continuum emission and
by emission-line fluxes. We therefore do not show results based on
this model here.

\subsection{Star formation and metallicity history parametrization}

In the past, the SFH of galaxies was typically modelled by an
exponential decay with a single parameter - for more complex models
one or two bursts of formation were allowed. In fact, it would be
better not to put any such constraints on star formation,
particularly considering that each galaxy may have (as a result of
mergers) several distinctly different aged populations. Star
formation takes place in giant molecular clouds, which have a
lifetime of around $10^7$ years. Splitting the history of the
Universe into the lifetimes of these clouds give a natural unit of
time for star formation analysis, but unfortunately it would require
several thousand of these units to map the age of galaxies formed 13
billion years ago, and the (lack of) sensitivity of the final
spectrum to the detailed history would make any estimate of star
formation history extremely degenerate.  We choose a compromise
solution, where we allow 11 time bins in which the star formation
rate (SFR) can vary independently. This allows a reasonable time
resolution, whilst not being prohibitively slow to compute.  For
most galaxies this parametrization is a little too ambitious, so we
do not recommend the use of the recovered star formation histories
on an individual galaxy basis.  Extensive testing \citep{PHJ03}
shows that for large samples the average star formation history is
recovered with good accuracy.  Future work will concentrate on
recovering only as much detail as the data from an individual galaxy
demands.   The boundaries between the 11 different bins used are
determined by considering bursts of star formation at the beginning
and end of each period (at a fixed metallicity) and set the
boundaries such that the fractional difference in the final spectrum
is the same for each bin.  This leads to a set of bins which are
almost equally spaced in log(lookback time).  Nine bins are spaced
with a ratio of log(lookback time) of 2.07 in this application of
MOPED, plus a pair of high-redshift bins to improve resolution at
$z>1$. This leads to a set of bins whose central ages are 0.0139,
0.0288, 0.0596, 0.123, 0.256, 0.529, 1.01, 2.27, 4.70, 8.50 and 12.0
Gyr. The gas which forms stars in each time bin is also allowed to
have a metallicity which can vary independently. The \cite{BC03}
models allow metallicities between
$0.02<\left(Z/Z_\odot\right)<1.5$. In order to investigate
metallicity evolution (Panter et al. 2007, in prep) no
regularization or other constraint is applied to the metallicity of
the populations - each different age can have whatever metallicity
fits best. A further complexity to the parametrization to deal with
post-merger galaxies which contain gas which has followed
dramatically different enrichment processes would be to have several
populations with the same age but independent metallicities.   It is
possible to consider a more complex parametrization, but again one
risks degeneracies in solution.  With 11 ages, 11 metallicities and
the dust parameter, the model has 23 parameters. The 23 dimensional
likelihood surface is explored by a Markov Chain Monte Carlo
technique outlined in \citet{PHJ03} Further information on the MOPED
algorithm is contained in \citet{Panter_thesis}.

\subsubsection{Speed issues}

MOPED \citep{HJL00} works by a massive data compression step,
forming (in this case) 23 linear combinations of the flux data.  The
resulting MOPED coefficients are fitted by standard minimum $\chi^2$
techniques. The data compression step is carefully designed to give
answers which are (in ideal circumstances) as accurate as performing
a full fit to the $\sim 3852$ flux data. One of the benefits of the
MOPED algorithm is that the number of compressed data is determined
by the number of model parameters, not the number of data points.
This has the huge advantage that analysis of the spectra at 3\AA\
resolution is no slower than analysis of 20\AA\ spectra (except for
a small increase in overheads such as pre-computing the data
compression weighting vectors).  The algorithm takes around two
minutes per galaxy on a fast desktop workstation, and represents a
speed up of around a factor of $170$ over a brute-force likelihood
fit.

\subsection{Computing Ensemble Results}

\begin{figure*}
\includegraphics[width=0.8\textwidth]{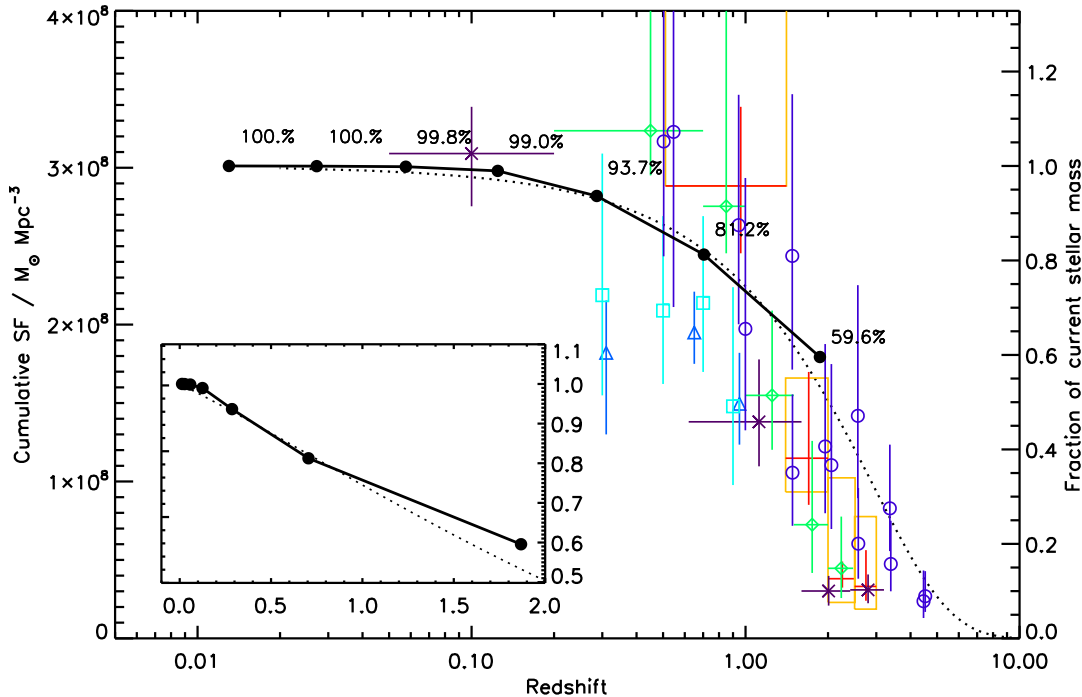}
\caption{The build up of the current stellar mass density of the
Universe, as determined from the SDSS DR3 galaxies and the recycling
fractions from the \citet{BC03} models with appropriate $V^{-1}_{\rm
max}$ weights (heavy dots, solid line). In all cases statistical
errors are contained within the dots marking the actual data points.
Also plotted for comparison are estimates from other surveys: HDF-N,
\citet{Dickinson03}(red plus, orange box denotes their error
estimation), HDF-S, \citet{Rudnick03}(purple stars), HST/CFRS,
\citet{Brinchmann00}(blue triangles), K20, \citet{Fontana04}(green
diamonds), FORS Deep and GOODS-South Fields, \citet{Drory05}(hollow
blue circles). For comparison with simulations we also show the
semi-analytical results based on the  Millennium Simulation of
 \citet{Croton06}(dotted line), normalised to our own final stellar
mass density. Insert shows exactly the same data plotted on a linear
redshift axis. No attempt has been made to convert between assumed
initial mass functions.} \label{fig:sf_build}
\end{figure*}

\subsubsection{Common star formation time bins}

MOPED determines the star formation history of each galaxy, relative
to the lookback time. To obtain the cosmic SFR it is necessary to
shift these results to a common set of time bins, which for
simplicity are chosen to be the same as those used in the galaxy
analysis. This allows direct comparison of galaxy star formation
between galaxies at different redshifts in terms of cosmic time, and
ensemble conclusions to be drawn. The shifting algorithm ensures
conservation of star formation. Star formation in the oldest bin is
always assigned to the final bin with ages no greater than 13.7 Gyr,
the age of the Universe determined by the concordant WMAP cosmology.
\begin{figure*}
\includegraphics[width=0.8\textwidth]{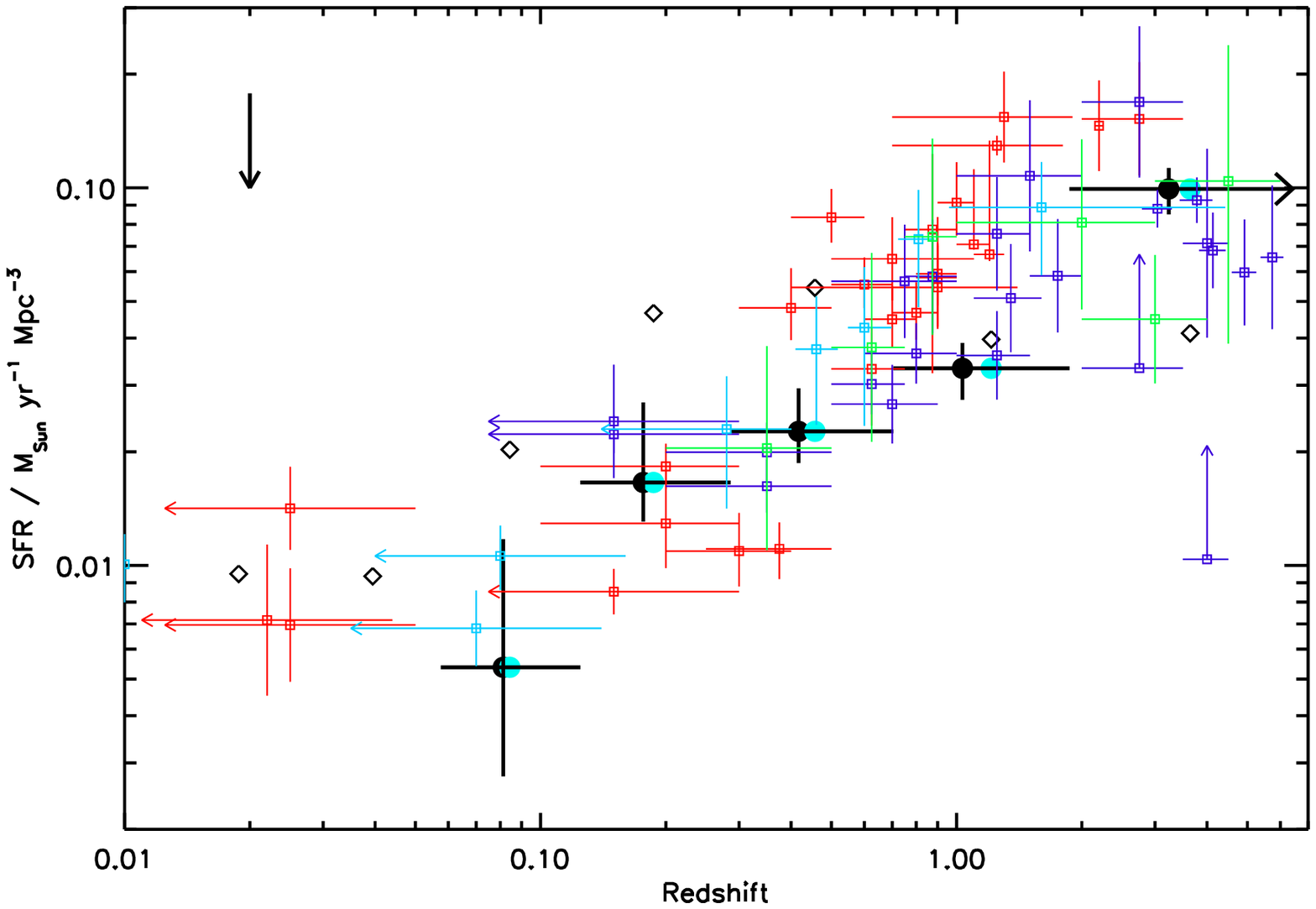}
\caption{The cosmic star formation history of the Universe, as
determined from the SDSS DR3 galaxies (heavy black horizontal bars
cover the bins, the luminosity weighted age of each bin given by
black dots). The vertical error bars are indicative of systematic
errors arising from choosing different subsets of the data for
analysis. See text for full details.  Also plotted for comparison
are the estimates from our DR1 work (black diamonds) \citep{HPJD04}
and those compiled in \citet{Hopkins04}, using the common
obscuration discussed within that paper. The points have also been
corrected for the minor difference in cosmology between that work
and this paper, and shifted downwards by 0.25 dex to convert from
the Salpeter to Chabrier IMF as indicated by the arrow in the top
left. The \citet{Hopkins04} points are coded as follows: UV
indicators (dark blue): \citet{Giavalisco04, Wilson02, Massarotti01,
Sullivan00, Steidel99, Cowie99, Treyer98, Connolly97, Lilly96,
Madau96}; OII, H-$\alpha$ and H-$\beta$ emission (red):
\citet{Teplitz03, Gallego02, Hogg98, Hammer97, Pettini98, Perez03,
Tresse02, Moorwood00, Hopkins00, Sullivan00, Glazebrook99, Yan99,
Tresse98, Gallego95};
 sub-mm (green)\citet{Flo:99, Barger00, Hughes98}; x-ray and radio (light blue):\citet{Condon02, Sadler02, Serjeant02,
Machalski00, Haarsma00, Con:89, Georgakakis03}.} \label{fig:sf}
\end{figure*}
\subsubsection{Inverse $V_{\rm max}$ weighting for Fossil studies}

In a magnitude-limited survey such as the SDSS the range of galaxy
types and sizes included in the survey will vary over the redshift
range studied. Some mechanism is required to compensate for this
change and determine the overall bulk parameters for the sample. To
convert from star formation rates to star formation rate density,
galaxies are weighted by $1/V_{\rm max}$, where $V_{\rm max}$ is the
maximum volume of the survey in which the galaxy could be observed
in the SDSS sample. This gives an unbiased estimate of the space
density $f$ of any additive property $F$ of the galaxy under
investigation, such as mass, luminosity, star formation rate.

\begin{equation}
f=\sum_{{\bf galaxies\ } i}\frac{F_i}{V_{max,i}}
\end{equation}
On smaller scales the estimator is affected by source clustering,
but the SDSS is deep enough that these variations should not be
significant. Any properties which change with redshift and which
could determine inclusion in the sample must be calculated for each
galaxy over the redshift range of the survey to determine whether or
not it would have been included. In order to calculate the $V_{\rm
max}$ assigned to each galaxy it is necessary to consider the
apparent magnitude and surface brightness evolution over the
redshift range. In order to compute this we use the same stellar
evolution models used in the MOPED analysis to calculate luminosity
over the lifetime of the galaxy due to its recovered star formation
history.

The magnitude of a galaxy over its lifetime depends on the
luminosity behaviour of the various stellar populations that make it
up - the star formation history. As young stars, the populations
will have a very high light output, which will reduce as they age.
This information is encoded in the galaxy spectrum and recovered by
MOPED, which gives the relative fractions of different aged
populations. To compute the observed magnitude if the galaxy were to
be observed at higher redshifts, we need to evolve the models over
time and track the changes in luminosity. Obviously, as the galaxy
is projected to a further redshift the younger fractions do not
contribute, as galaxy is being `observed' before these populations
were born. Since the spectral energy distribution of the light
changes with evolution of the galaxy, it is also necessary to apply
the filters used by Sloan to determine the flux included in the $r$
band, leading to a change in $c_{i,z}$ in the $r$-band magnitude as
the redshift is changed:

\begin{equation}
r_{i,z}=r_{i,z_{\rm obs}}+c_{i,z}.
\end{equation}

These magnitude corrections can then be used to calculate
corrections which need to be applied to the surface brightness. The
surface brightness of the galaxy at each redshift $z$ is

\begin{equation}
\mu_{i,z}= r_{i,z} + c_{i,z} + 2.5\log_{10}\left[\pi r_{50,i}^2
(D_{z}/D_{z_{obs}})^2\right] + 2.5\log_{10}2
\end{equation}

where $D_z$ is the luminosity distance, $r_{50}$ is the Petrosian
half light radius and $z_{obs}$ is the observed redshift of the
galaxy. This equation assumes that the size of the galaxy does not
change over the redshift range. Although this assumption is valid
for low redshift sources, it will need to be developed if the
technique is applied to deeper surveys.

The MOPED technique gives the relative strengths of the different
spectral models. The mass originally created to make these masses is
then calculated, and by dividing this by the maximum volume over
which the galaxy could be observed gives the star forming density,
$\rho_{i}$. By adding all the $V_{\rm max}^{-1}$ weighted star
forming densities of galaxies in the sample, rebinned to a common
time frame, the overall star forming density $\rho$ can be found for
the region studied.

Galaxies may only contribute to the SFR of a time bin if they are at
a lower redshift than the lower limit of that bin. This conservative
approach ensures that the star formation of a galaxy is never
extrapolated. It also means that in all but a few cases where a
galaxy is almost on the boundary between bins, the youngest
populations, those which are better calculated using instantaneous
indicators from emission lines, do not contribute to estimates of
the SFR. In addition, to ensure that our results are not biased by
single erroneous SFH reconstructions we only present results from
bins which contain contributions from greater than $1000$ galaxies.

\section{Results from the DR3}

\subsection{Evolution of Stellar Mass Density}

The simplest interpretation of the combined star formation histories
is a simple $V^{-1}_{\rm max}$ weighted addition that shows how the
present day stellar mass of the Universe has built up. In Fig.
\ref{fig:sf_build} and Table \ref{tab:SFB} we show how the stellar
mass density of the universe has changed since $z=2$. In order to
compute the remaining mass (when from the MOPED algorithm we
estimate the original mass of stars formed) it is necessary to
invoke the recycling fraction, R. Rather than assume a blanket
correction, we have used the detailed predictions allowed by the
\citet{BC03} models for each population of each galaxy, based on
their determined ages and metallicities. Although this is slightly
more laborious than assuming a blanket recycling fraction it gives a
more accurate determination of present mass, since R is a function
of both age and metallicity.

The abscissa on this plot refers to the minimum redshift of a given
bin, when all the stars formed in that bin will be in place. Simple
interpolation between these points is valid as long as the star
formation rate is constant across the bin. From this it is easy to
see that approximately 60\% of the present day stellar mass was in
place by $z=1$, and 90\% by $z=0.35$. It is also clear that in our
sample there is overall very little star formation at $z<0.1$.

\begin{table}
 \centering
 \begin{minipage}{\columnwidth}
  \caption{The buildup of stellar density derived from the MOPED/SDSS-DR3 fossil record.}

\begin{tabular}{|c|c|c|c|}
  \hline
  $z$ & Stellar Mass Density & Min Density & Max Density\\
   &  & log$_{10}$(M$_\odot$Mpc$^{-3}$) & \\
  \hline
 0.0130  &     8.459  &     8.401 & 8.558\\
 0.0272   &    8.459     &  8.401 & 8.557\\
 0.0575    &   8.458  &     8.401&  8.556\\
 0.125  &     8.454 &      8.399 & 8.549\\
 0.286   &    8.427   &    8.377  & 8.516\\
 0.705    &   8.362    &   8.317 & 8.447\\
 1.87     &  8.222     &  8.186  & 8.309\\
  \hline
  \label{tab:SFB}
\end{tabular}
\end{minipage}
\end{table}

For comparison we also plot the mass buildup estimated from various
surveys. We also show the results from semi-analytical modelling
based on the Millennium Simulation \citep{Croton06}. The data shows
excellent agreement between the largest astronomical survey and the
largest simulation ever undertaken. The statistical errors on the
points in Fig.{\ref{fig:sf_build}} are so small that they are
contained within the points. This is of course unrealistic, and we
have also included a more realistic treatment of errors in our
ensemble estimates by selecting many different (but still
reasonable) sample criteria and comparing the ensemble results that
result from each sample. These separate samples vary between minimum
redshift (0.001, 0.003, 0.006, 0.009, 0.01, 0.03, 0.05), maximum
redshift (0.2, 0.3, 0.34), bright limiting magnitude (13.5, 14.0,
14.5, 15.0), faint limiting magnitude (17.7, 17.77) and surface
brightness cuts (23.0, 23.5, 24.0). They show the maximum and
minimum recovered quantities from the many ways of selecting source
galaxies, and should be indicative of the systematic errors from
basing the analysis on different galaxies, although some of this
variation will come from real differences between the histories of
the galaxies selected. The extents of these systematic (and heavily
correlated) error approximations are not plotted, but are given in
Table \ref{tab:SFB} as maximum and minimum densities. In the case of
the cumulative stellar mass density the overall mass can vary by up
to 25\%.

\subsection{The cosmic star formation rate}

In Fig. \ref{fig:sf} we show the cosmic star formation recovered
from SDSS DR3 with a Chabrier initial mass function, using the
Bruzual \& Charlot (2003) 3\AA\ resolution spectral synthesis
models, and a single-parameter dust screen following the extinction
curve of the LMC.

As seen before in many studies, the steep decline in SFR is clearly
demonstrated. In contrast to our previous study, we do not find a
peak in SFR at $z<1$.  This was also found by \citet{Mathis06}, in
work based on the MOPED fossil analysis approach. If there is a
peak, it occurs somewhere in our last bin ($z \gs 2$). These new
results from the fossil record are in much better agreement with
determinations based on contemporary star formation rates. Purely
statistical error bars are so small to be almost invisible in all
but the lowest redshift bin. As with the previous figure, the
vertical error bars in Fig.{\ref{fig:sf}} are indicative only.  They
show the maximum and minimum recovered star formation rates from the
many ways of selecting source galaxies. They should be indicative of
the systematic errors from basing the analysis on different
galaxies, although some of this variation will come from real
differences between the histories of the galaxies selected.

Although for completeness we show our results from $z \sim 0.1$, we
do not present points with $z \ll 0.1$ for three reasons. First,
from Fig. \ref{fig:sf_build} it is clear that there is much less
mass in these bins on which to form an estimate of the SFR, the
change in mass with time. Second, to avoid biasing, our $V_{\rm
max}$ criteria exclude galaxies from contributing to bins with upper
boundary lower than their redshift - hence the bulk of the galaxies,
at approximately $z=0.1$, can only contribute to bins from $z=0.2$
onwards. Third, the galaxies contained in these bins, and the
resultant SFR, are strongly dependent on sample criteria, as
expected when sample size drops. Later in this paper we analyse a
subset of the data to determine which changes are responsible for
the modifications to the results; the main reasons for the changes
are the better calibration of the SDSS DR3, the change in IMF to
\citet{Chabrier} and the change to the higher-resolution Bruzual and
Charlot (2003) models. The star formation rate density resulting
from the MOPED DR3 analysis is presented in table \ref{tab:SFR}.

\begin{table}
 \centering
 \begin{minipage}{\columnwidth}
  \caption{The SFR density derived from the MOPED/SDSS-DR3 fossil record.}

\begin{tabular}{|c|c|c|c|c|c|}

  \hline
  $z$ &  $z_{min}$ & $z_{max}$ & SFRd  & Min SFRd & Max SFRd\\
   &  & &  & (M$_\odot $yr$^{-1}$Mpc$^{-3}$)&\\
  \hline
0.081 & 0.0575 & 0.125 & 0.00537 & 0.00276 & 0.0117 \\

0.177 & 0.125 & 0.286 & 0.0166 & 0.0131 & 0.0270  \\

0.416 & 0.286 & 0.705 & 0.0226 & 0.0187 & 0.0294  \\

1.03 & 0.705 & 1.87 & 0.0332 & 0.0275 & 0.0388  \\

3.24  & 1.87 & 6.42 & 0.0993 & 0.0850 & 0.113 \\
  \hline
%\caption{Star Formation Rate Density from MOPED Analysis}
  \label{tab:SFR}
\end{tabular}
\end{minipage}
\end{table}

\subsection{The mass function of stellar mass and $\Omega_{b*}$}

\begin{figure*}
\includegraphics[width=0.7\textwidth]{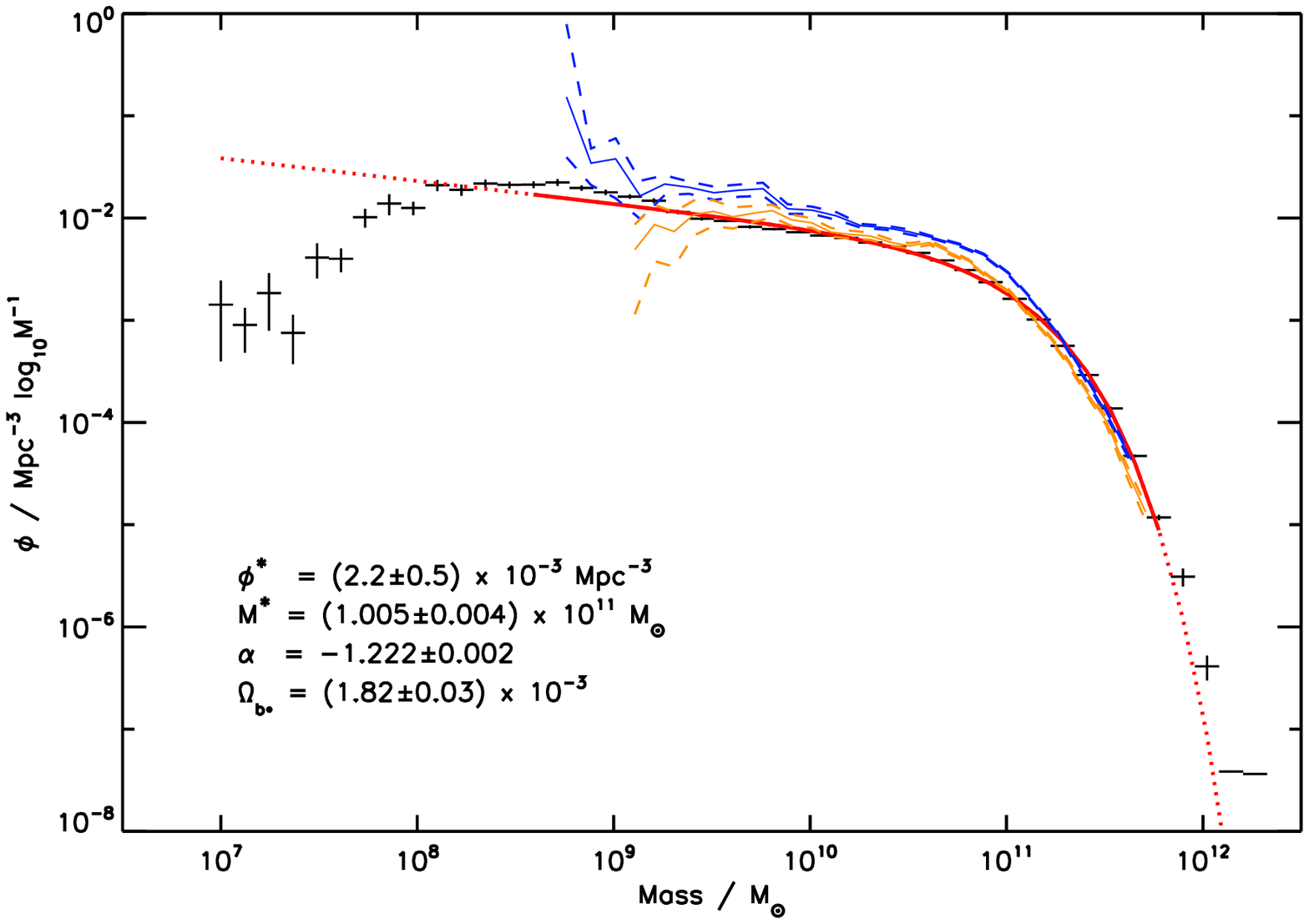}
\caption{The galaxy stellar mass function of SDSS.  Also shown are
those of \citet{Cole01} (Orange) and \citet{Bell03SMF} (Blue),
corrected for differences in IMF. The errors quoted on the
parameters are based upon the statistical errors of the galaxy
sample rather than systematics, which are discussed in the text. The
red solid line is the \citet{Schechter76} best fit solution over the
qualifying points, the dotted section is an extrapolation over the
range covered by the data.} \label{fig:massfn}
\end{figure*}

The galaxy stellar mass function of SDSS DR3 is shown in Fig.
\ref{fig:massfn}, for a range of almost 5 decades in mass
($10^7-10^{12}\,$M$_\odot$). The errors shown are statistical based
on our chosen sample criteria. We also compute the mass function for
the alternative sample criteria to develop systematic errors. Over
much of this range a Schechter fit is good, with parameters
$\phi^*=(2.2\pm0.5_{stat}\pm1_{sys}) \times 10^{-3}$\ Mpc $^{-3}$,
$M^*=(1.005\pm0.004_{stat}\pm0.200_{sys}) \times 10^{11}$ M$_\odot$,
and slope $\alpha=-1.222\pm0.002_{stat}\pm0.1_{sys}$ calculated in
the region where there are more than 300 galaxies contributing to
each bin ($10^{8.5}-10^{11.85} M_\odot$). The mass function is very
similar to our DR1 analysis \citep{PHJ04}, but shifted to lower
masses as a result of the use of the Chabrier IMF rather than
Salpeter. For further discussion of the systematic differences in
recovered galaxy mass caused by IMF variation refer to the
discussion in \citet{Bell01}.

The stellar mass function can be used to give a further constraint
on the contribution to the density parameter from baryons in stars,
$\Omega_{\textrm{b*}}$. By integrating the mass over the range of
the mass function we deduce a value of $\Omega_{\rm b*}=(1.82 \pm
0.03_{stat} \pm 0.1_{sys} )
 \times 10^{-3}$ (systematic error). This
value is in broad agreement with results obtained previously when
the correction from Salpeter IMF is taken into account
\citep{PHJ04,Cole01,Bell03SMF,Fukugita98,Kochanek01,Glazebrook03,Persic92,Salucci99}.
The statistical errors reflect the spread of results obtained with
varying sample criteria as before.

\subsection{Downsizing}

\begin{figure*}
\includegraphics[width=0.7\textwidth]{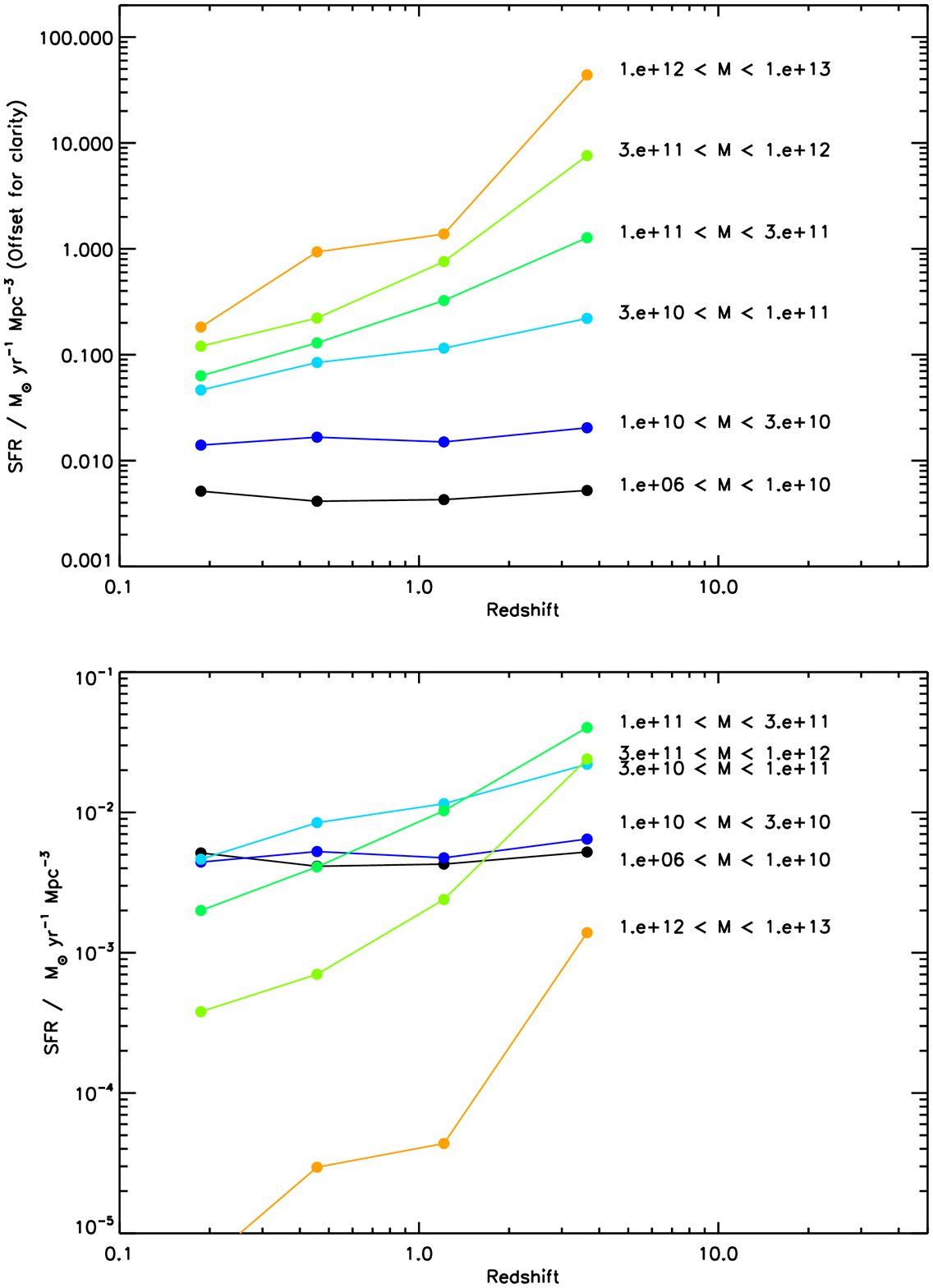}
\caption{The star formation rate of galaxies of different masses.
These plots show the contribution to the overall star formation rate
in the universe from galaxies with different masses over the
redshift range we consider reliable. In the upper panel the SFR has
been offset to enable easier comparison of the curves, in the lower
there has been no offset applied. It is clear that the more massive
systems formed their stars earlier, although no conclusion can be
drawn as to the number of objects these stars formed in.}
\label{fig:downsizing}
\end{figure*}

One of the results of \citet{HPJD04} was the finding of `downsizing'
from the SDSS fossil record. Using the new models at higher
resolution we have found that the evidence for downsizing is just as
clear. In Fig. \ref{fig:downsizing} we show the cosmic star
formation rate for galaxies split into different stellar mass
ranges. A clear signature of `downsizing' is seen: the stars ending
up in today's highest-mass galaxies formed early, and show
negligible recent star formation, while the lower-mass galaxies
continue with star formation until the present day. The lower,
non-offset plot can be used to determine for a given redshift which
galaxies dominate the star formation rate.

% \subsection{Dust content of galaxies}
%
%Fig. \ref{fig:dust_conc} shows the estimates of the present day dust
%E(B-V) parameter for the galaxies, and their dependence on the
%concentration index $c$.  The values in the lower-right are
%essentially non-detections, so we see that the high concentration
%(ellipticals) are largely devoid of dust, whereas the low
%concentration (spirals) have clear dust detections in almost all
%cases.  The nominal cut between the morphological types is usually
%taken to be $c=2.86$ \citep{Shen+03}, and with further verification
%could be taken as a spectroscopic morphology proxy for higher
%redshift galaxies in the survey where imaging is not possible.
%
%\begin{figure}
%\includegraphics[width=\columnwidth]{latest_dust_conc.ps}
%\caption{A 2D histogram showing the joint distribution of dust
%content and $r$ band concentration index for all galaxies in the
%Sloan Main Galaxy Catalogue. The intensity is in proportion to the
%number of galaxies in each cell. The dust provides a neater split
%into early- and late-type galaxies than the concentration index,
%allowing a spectroscopic determination of galaxy type if the
%spectrophotometry is reliable. Note that the low dust values in the
%peak in the lower right should be regarded as upper limits.}
%\label{fig:dust_conc}
%\end{figure}
%

\subsection{Investigating spectral residuals}

\begin{figure*}
\includegraphics[width=15.5cm]{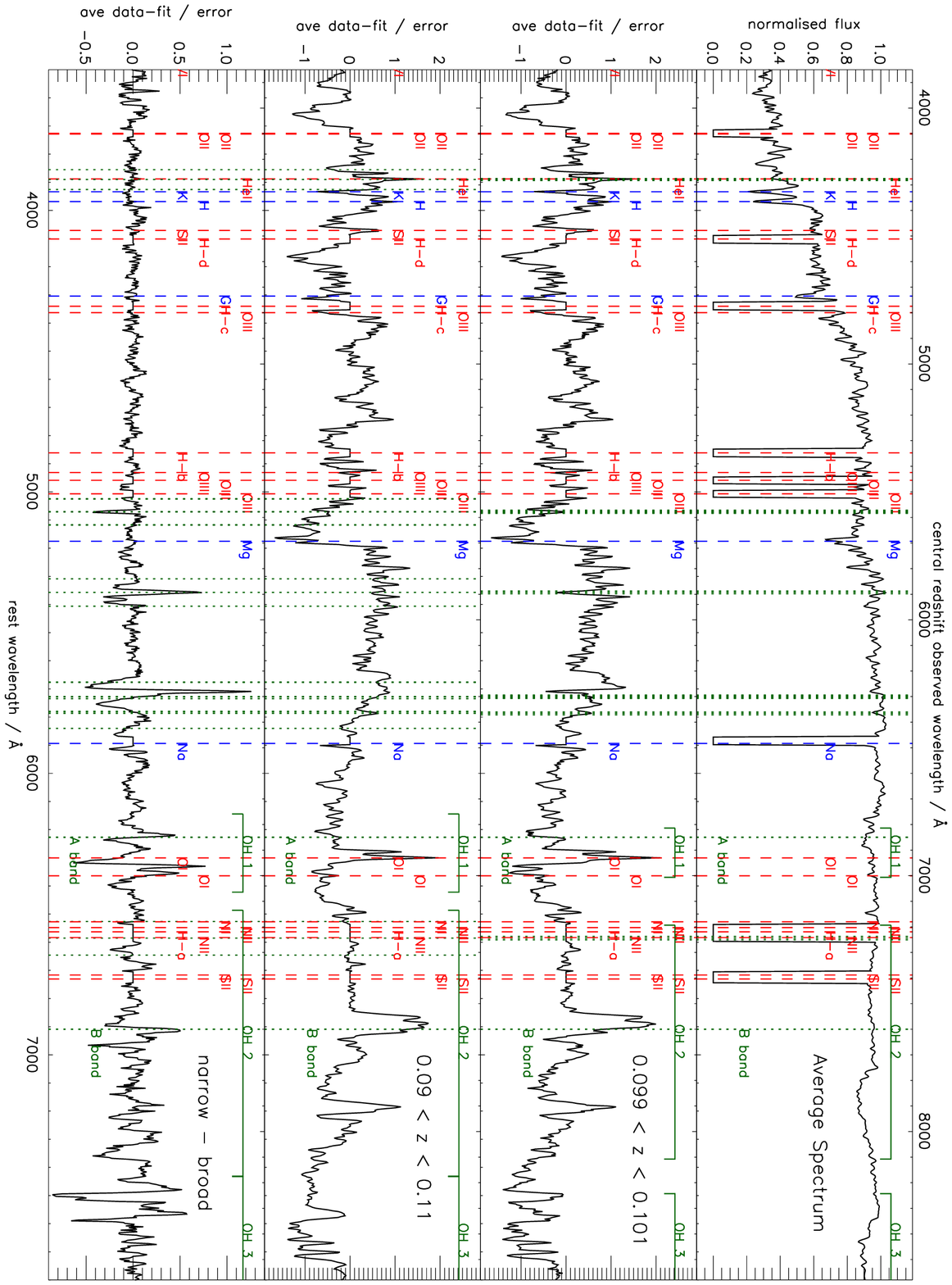}
\caption{The mean spectrum and the mean of the residual divided by
the error for MOPED fits of DR3 galaxy spectra. The first panel
shows the average spectrum for the galaxies in with high S/N and
$z=0.1\pm0.01$. The second and third panels show the residual at
each pixel divided by the error at that pixel {\it averaged in the
rest frame}. The first plot shows the average for 747 high S/N
galaxies within the range $z=0.1\pm0.001$, the second for 7406 high
S/N galaxies within $z=0.1\pm0.01$. The third plot shows the first
residual set minus the second, and clearly separates areas
contaminated from skylines (which are broadened in the second plot
by the redshift range). Emission features are labelled in red,
absorption blue and skylines green.} \label{fig:dr3_resids}
\end{figure*}

Due to the power of MOPED and the number of SSPs offered for fitting
(11) excellent fits can be obtained if the models are accurate. By
comparing the residuals of the best fitting spectrum to the raw data
on a pixel by pixel basis and then averaging over many galaxies
\textit{in the galaxy restframe} it is possible to determine exactly
which areas are not being accurately fitted for a given spectrum. By
stacking the residuals of high signal to noise (Sloan {\sc
specobjall.sci\_sn}, science $S/N > 20$ per flux measurement)
galaxies it is possible to determine to a high degree of accuracy
which wavelength ranges are failing in the models, and to what
extent. If we assume that MOPED can, in most cases, obtain the best
possible fit to a spectrum then the differences must be features not
included in the models. These features could be things that the
models are not designed to measure (instrumental effects,
interstellar or intergalactic medium absorption, skyline
contamination etc.) or alternatively features that are either
misrepresented or not yet included in stellar modelling codes (e.g.
alpha enhancement \citep{Thomas03}, emission lines, helium
production of $dY/dZ \sim 1.5 - 2$ \citep{Jimenez03} and other
spectral features). Fig \ref{fig:dr3_resids} provides some insight
into which features in the residuals can be identified. Since we
subtract the modelled spectrum from the data, unfitted emission will
have a positive effect on the residuals while unfitted absorption
will be negative. Both obvious emission lines and filling of
absorption features should be detectable. The spectra used for this
analysis are those which have already had the strongest emission
line regions removed, as detailed earlier - in this case the
residual is simply zero.

The first panel of Fig. \ref{fig:dr3_resids} shows the mean
spectrum. To distinguish between galaxy features and skyline
features we select two redshift ranges with the same central
redshift, $z=0.1$. Since the averaging of residuals is carried out
in the galaxy restframe, increasing the redshift spread will act to
spread any skyline features. The third panel gives the residuals for
galaxies within 0.001 of the central redshift while the second has a
range of 0.01. The fourth panel shows the second subtracted from the
first. In this case, skyline/instrumental will create regions with
large amplitude. The relevant skyline features (and their
convolutions with the extremes of the particular redshift
distribution) are shown and correspond exactly. It can be seen from
the third panel that the majority of the spectral range is
remarkably clear of skyline contamination - testament to the high
quality of the Sloan spectroscopy.

Considering the regions which are not excluded by skylines we begin
to be able to assess the ability of the \citet{BC03} models to fit
the data. It is clear that in all regions where weak emission
filling could be present and has not been masked there is a slight
positive tendency in the residuals (although this is of course not a
failing of the models but a reflection of possible systematics
affecting our fits), and many of the features in the residual
correspond to features that may be affected by alpha enhancement. A
more detailed study of individual indices and their alpha
enhancement, and their resolution in more advanced models including
variable abundance ratios will be presented in a later paper.
Although the vast majority of features in the residual can be
directly related to specific lines there is a strong signal just red
of the magnesium line at restframe 5176\AA. Although tempting to
attribute this to poor fitting of the magnesium feature, it could
also be interpreted as part of a broader feature between restframe
5200-5900\AA. It is interesting to compare the fitting in this
region to the area around the 4000\AA\ break as both are heavily
dependent on metallicity. It is surprising that the models do so
well in fitting the break but cannot simultaneously fit this region,
and taken with the excess around the Calcium Hydride band
(~6830-6900\AA) suggest that perhaps the balance of K-M giants in
the models must be improved, the resolution of which would lead to a
bluer continuum.

Converting this feature to the observed frame it coincides with the
dichroic region of the combined red and blue SDSS spectrographs. If
this were to be the cause one would expect that the feature would
appear in the third panel, which it does not, and certainly further
work would be necessary to interpret this residual feature in the
context of the DR3 spectrophotometric calibration pipeline.

\section{The Impact of Model Choice}

In this section, we investigate the influence on assumptions on the
results obtained in the last section.  There have been several
improvements and changes since our analysis of DR1, and the results
have changed to some degree.  The purpose of this study is to see
how the assumptions change the conclusions, and to get some idea of
the systematic effects introduced by choices of such things as the
stellar populations used to model the spectra.

\subsection{Sample studied}

Although the MOPED algorithm allows rapid analysis of different
modelling choices, to investigate a wide range of parameters it is
necessary to cut the sample down to something more manageable. We
chose to operate on the main galaxy sample spectra in plates 0288
and 0444. These two plates form a representative sample of 808
galaxies from two widely separated patches on the sky. In total 767
of the 808 satisfy the criteria used in our main analysis and would
appear in our estimates of SFR.

\begin{figure*}
\includegraphics[width=15cm]{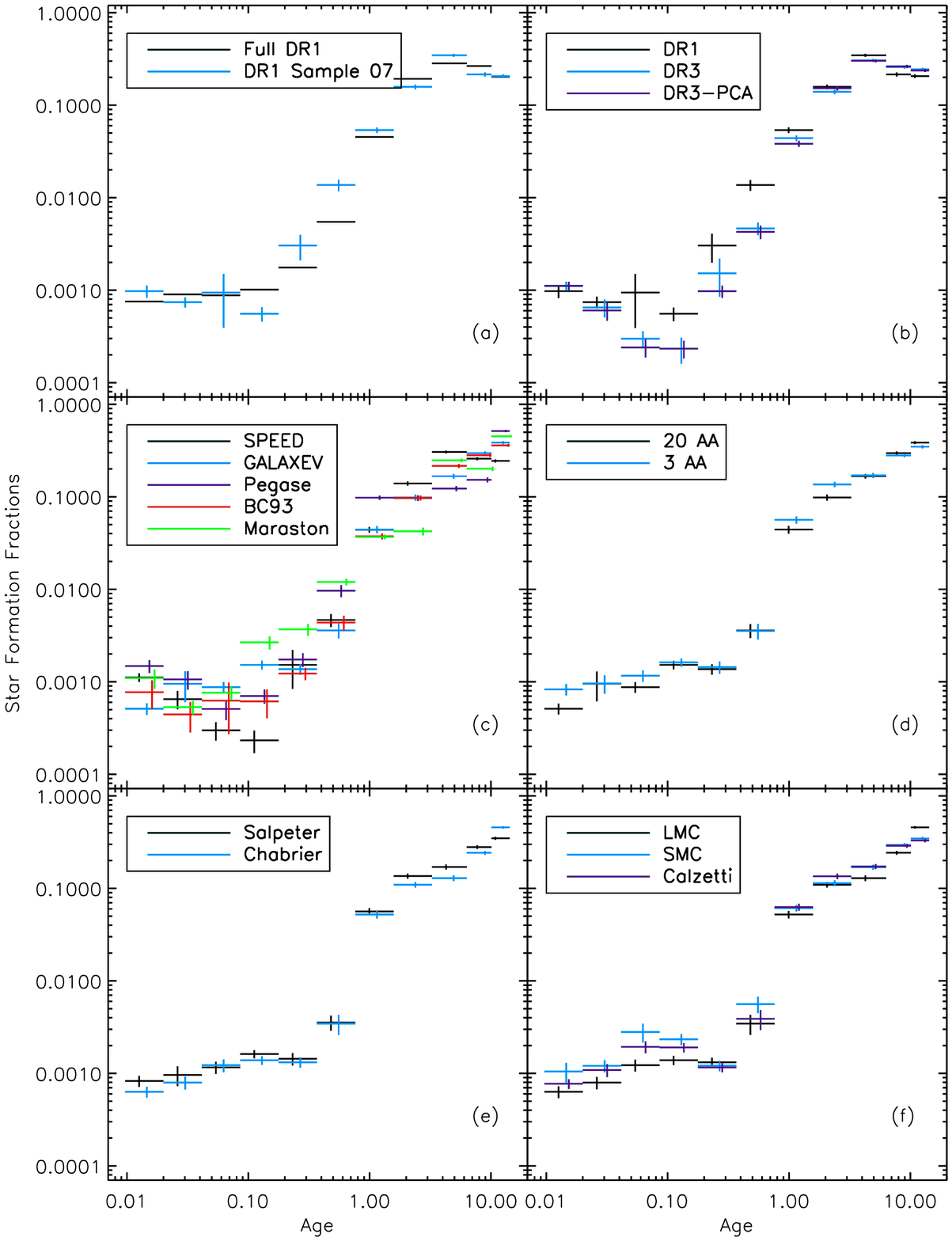}
\caption{The systematic changes in recovered SFH for various
modelling choices. The labels show the various modelling choices
applied to generate the SFFs, and are expanded upon in the text. It
is important to note that these are the fractions of total mass
formed over the lifetime of the galaxy, and not the fractions of
light contributing to the recovered spectrum. The light from the
youngest populations is some 300x brighter than the oldest. The
fractions are normalised, so changes in overall recovered mass for
spectra are not apparent.} \label{fig:sff}
\end{figure*}

\begin{figure*}
\includegraphics[width=16cm]{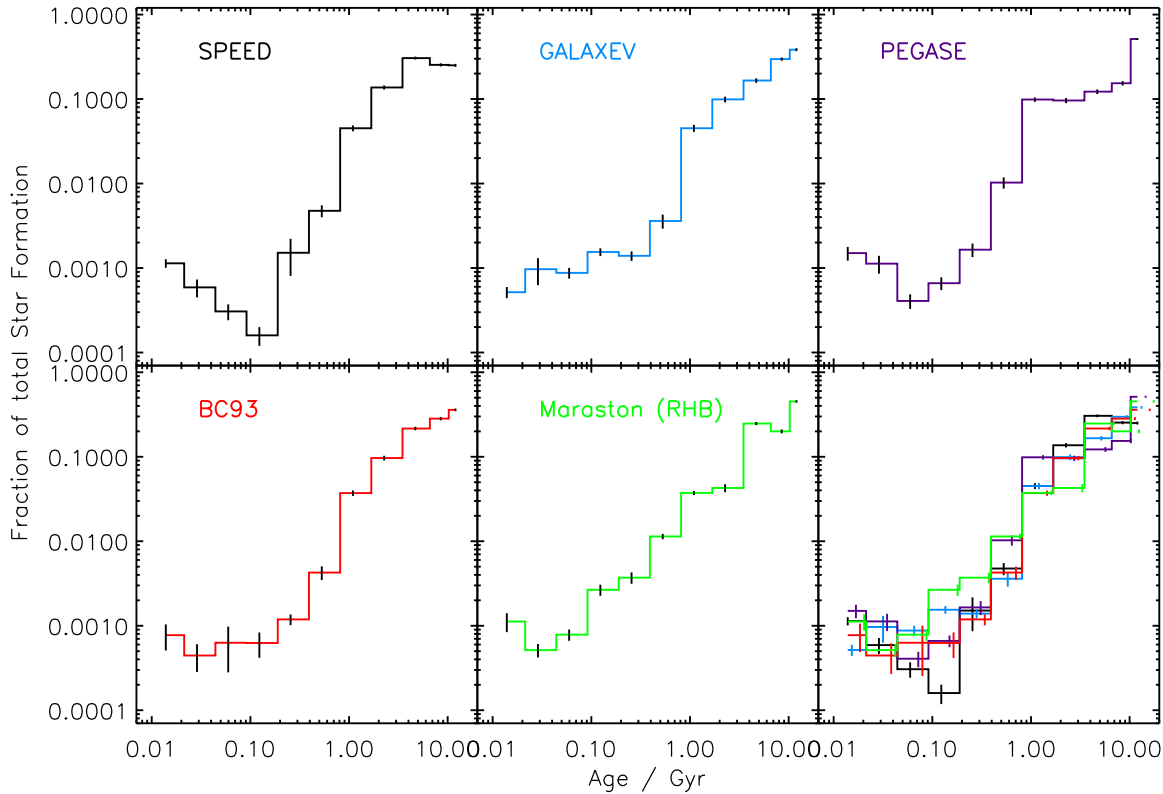}
\caption{The systematic changes in recovered rest frame star
formation fractions (not rates) for various different SSPs expanded
from figure \ref{fig:sff} for clarity. It is important to note that
these are the fractions of total mass formed over the lifetime of
the galaxy, and not the fractions of light contributing to the
recovered spectrum. The light from the youngest populations is some
300x brighter than the oldest} \label{fig:ssp}
\end{figure*}

\begin{figure*}
\includegraphics[width=15cm]{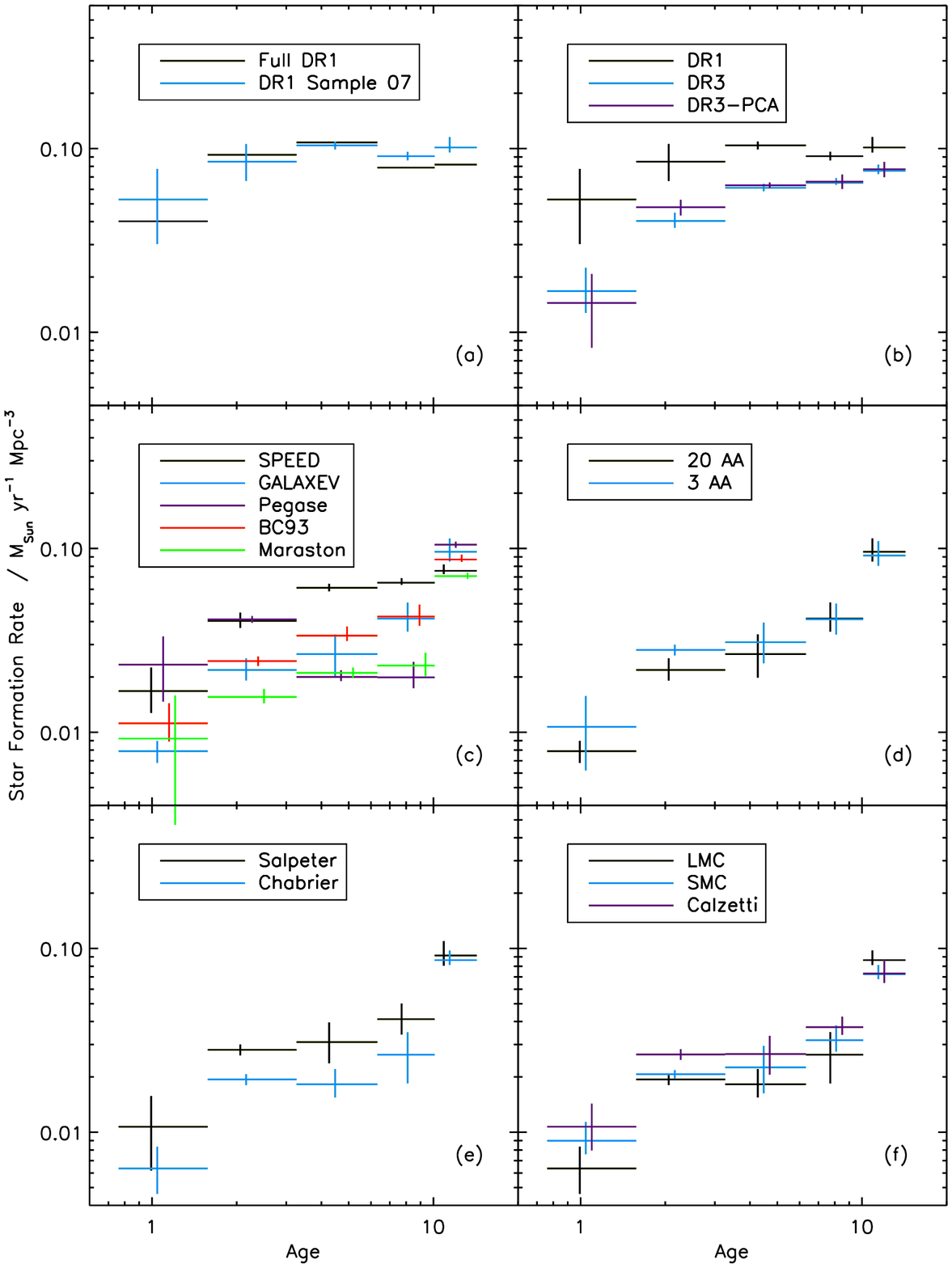}
\caption{The systematic changes in the overall SFR for various
modelling choices. The labels show the various modelling choices
applied to generate the SFFs, $V_{max}$, mass and SFR and are
expanded upon in the text. We plot only those bins used for the
analysis in the main section, and caution that due to the large
$V_{max}$ corrections there are insufficient galaxies in this
sub-sample to draw robust conclusions} \label{fig:sfr_var}
\end{figure*}

\subsection{Star Formation Fractions and Star Formation Rates}

There are two places where modelling choices affect the shape of the
recovered cosmic star formation rate - the estimation of the
relative fractions of different SSPs and the calculation of a
$V_{\rm max}$ correction based on that SFH. For this analysis we
wish to investigate the two stages independently. To achieve this we
need to compare both the SFF recovered using different models and
the SFR that results.

The galaxies' star formation fractions (SFF) and total mass are
calculated in their individual rest frames. This SFH is then
converted to a common frame for all galaxies and weighted in
proportion to $1/V_{\rm max}$. Both the original SFF and the
$1/V_{\rm max}$ weighting are calculated by the models, and the mass
is dependent on the IMF.  Our first comparison (Fig. \ref{fig:sff})
shows the initial normalized SFF recovered for the galaxies
\textit{averaged in their restframes}. Since it is our intention to
compare the relative fractions produced by changing model parameters
rather than the overall mass (which has been described elsewhere,
see \citet{Bell01}) we present the SFF normalized to total star
formation of 1 per galaxy . This should give a clear indication of
differences on recovered star formation \textit{fractions} in the
models.

Next we compare the cosmic SFR calculated using these various
modelling choices. This method allows direct comparison of the SFR
obtained using each model, but is subject to greater errors due to
the sensitivity of the $V_{max}$ calculation on individual star
formation histories and the large relative weight change this can
produce. We present this model-dependent SFR in addition to the SFF
but caution that the number of galaxies present is insufficient to
draw robust conclusions. Given the substantially smaller number of
galaxies, the errors are far larger than those of the complete DR3
analysis presented earlier in this paper.

 A summary of the models is given in table \ref{tab:models};
unless otherwise stated a \citet{Salpeter} IMF is used to calculate
the SSPs.

\begin{table}
 \centering
 \begin{minipage}{\columnwidth}
  \caption{A summary of models used in the analysis.}
\begin{tabular}{|c|c|c|}
  \hline
  Name & Reference & FWHM \\
  \hline
  SPEED & \citet{JMDPP04} & 20\AA \\
  PEGASE & \citet{Pegase} & 20\AA \\
  BC93 & \citet{BC93} & 20\AA \\
  Maraston & \citet{Maraston05} & 20\AA \\
  GALAXEV & \citet{BC03} & 3\AA \\
  \hline
  \label{tab:models}
\end{tabular}
\end{minipage}
\end{table}

\subsection{Sample galaxies compared to full dataset}
Fig {\ref{fig:sff}}a shows the difference in recovered SFF between
our original DR1 analysis and our subsample. The differences are due
to the number of galaxies (120,000 in the full DR1 vs. 808 in the
DR1 Sample). For all further comparisons in this section the sample
of galaxies will be 808 identified in the MGS plates 0288 and 0444.
Fig {\ref{fig:sfr_var}}a shows the difference in SFR.

\subsection{Pipeline changes DR3 vs DR1, PCA cleaning of skylines}
Fig {\ref{fig:sff}}b shows the impact of changing the pipeline used
to reduce the original raw data. DR3 contains a more accurate
calibration of the continuum of the spectrum and a systematic trend
that deviated the continuum blueward of $4000$ \AA\ has been
corrected. The SPEED model, at a resolution of $20$\AA\, was used
for all three sets. Except for bins 6-9 (numbered from the right),
the variation the two datasets introduce is very small, at the few
percent level. However, for these bins the deviations are as much as
a factor of three. Note that, where present, continuum discrepancies
between the blue end for DR1 and DR3 were on average 10\% in the
flux (at 3500 \AA). It is remarkable that such drastic changes
introduce such small changes on the average physical properties of
the sample.

It is reassuring that the changes made to the spectra by PCA removal
of skyline regions have virtually no effect on our SSP fits. It is
important to realise though that the majority of the regions cleaned
by the \citet{Wild05} code are outside of our sampled wavelength
range.

The effect of the pipeline on the SFR shown in Fig
{\ref{fig:sfr_var}}b is far greater. The mean flux of the spectra
between the two releases has changed by a factor of between 1.5 and
2, and the masses reflect this change. Even ignoring this change in
the normalization of the SFR it is clear that the trend for a low
redshift peak is no longer present - the SFR appears to fall
monotonically to the present day. This interpretation does not
exclude a peak at a higher redshift than our eldest bin.

\subsection{Stellar population models} Figs. {\ref{fig:sff}}c and {\ref{fig:ssp}} show the comparison between five
different stellar population synthesis models: the \citet{JMDPP04}
SPEED models, the \citet{Pegase} PEGASE models, the
\citet{Maraston05} RHB models; the \citet{BC93} models and the more
modern 3\AA\ \citet{BC03} GALAXEV models rebinned to 20\AA. The
comparison is done at 20\AA\, for the Salpeter IMF and for the
one-parameter dust model. It is important to establish that this
analysis cannot say which model set is `right', only assist in
understanding the differences between models. The overall shape of
the SFF is in reasonable agreement - although there are certainly
discrepancies between the populations that are recovered. For the
very oldest populations the different models agree very well. This
is not entirely unexpected of course, as the stars which contribute
to this area of the age-metallicity parameter space are well studied
and dominate the emission at red wavelengths. The different models
also predict roughly similar proportions of the very youngest
populations, which rely on similar prescriptions for the evolution
of blue massive stars and can be constrained through the emission at
the bluest wavelengths. At intermediate ages the agreement is not so
good: this is likely to be caused, at least in part, by the
difficulty in recovering the fraction of intermediate-age stars in
stellar populations with declining star formation histories (see
discussion by \citet{Mathis06}). This tends to produce an artificial
step around 1 Gyr in the star formation history, except perhaps in
the \citet{Maraston05} model. The different behavior of this model
probably results from the different prescription for bright
Thermally Pulsing Asymptotic Giant Branch (TP-AGB) stars. The
contribution by these stars to the integrated light is still subject
to controversy in current population synthesis models. Since our
spectra are fitted in addition, any poorly fit component will be
replaced by another.

Fig. {\ref{fig:sfr_var}}c shows the difference that the various
models make to the recovered SFR. It is clear that there is a large
spread - as discussed earlier, this is due to the fact that the
models contribute both to the estimation of the SFH and the $V_{\rm
max}$ weights attributed to the galaxies. In all models except those
of the PEGASE group the SFR decreases monotonically from the oldest
to the youngest bins.

\subsection{The impact of resolution, 20\AA\ vs 3\AA}
Fig.{\ref{fig:sff}}d shows the impact of increasing the spectral
resolution of the data. In this case we use the BC03 at resolutions
of 20\AA\, and 3\AA\, on DR3. The differences between the two curves
are very small. the maximum deviation is only of about 30\% in bin 4
and smaller in bin 5. For the other bins the agreement is
remarkable. What this comparison is telling us is that 20\AA\ is
sufficient resolution to determine the average properties of
galaxies. The higher resolution does not add much extra information
to this. More importantly, the result is not biased at the lower
resolution. This is not entirely surprising since the continuum
certainly contains information about both age and metallicity of a
stellar population (e.g. \citet{JMDPP04}). The shape of the SFR
recovered in Fig. {\ref{fig:sfr_var}}d is remarkably consistent
between the two resolutions, with a significant variation in only
one bin.

\subsection{IMF: Salpeter vs. Chabrier.}
Fig.{\ref{fig:sff}}e and {\ref{fig:sfr_var}}e show the impact of
changing the initial mass function on the recovered SFF and SFR. In
this case we have chosen to work with the BC03 models at 3\AA\,
resolution. The IMF determines the initial distribution of the
number of stars as a function of mass. It is therefore not
surprising to find changes in the recovered star formation history
for dramatic changes in the IMF. The Chabrier and Salpeter IMF are
very similar for masses larger than $1-2$ M$_{\odot}$ (they are both
power laws) while they differ considerably at smaller masses: the
Salpeter IMF continues being a power law (with index $-1.35$) while
the Chabrier IMF deviates containing a much smaller number of small
mass stars relative to the Salpeter IMF. As panel {\ref{fig:sff}}e
shows the main difference occurs at the oldest bin (bin 1), with
smaller differences in bins $2-4$ and virtually no differences for
the youngest bins when the bootstrap error bars are taken into
account. This effect is propagated through the $V_{\rm max}$ and
mass calculation to the SFR. This is more-or-less what one would
expect: for the oldest bin, the Chabrier IMF has to compensate its
relative lack of low mass stars compared with the Salpeter IMF by
forming more of them. The differences then disappear as more massive
stars are more dominant at recent ages. It is interesting that
although our results will be affected by varying the IMF, it will be
in a different sense from the results from instantaneous SFRs
attained at high redshift. Where as virtually all instantaneous
indicators estimate the total mass of stars from the very high mass
UV emitting stars, our technique uses the low mass remnants. The
fact that the two approaches agree suggests that the IMFs currently
in vogue are along the right lines, and a direct comparison of
sufficiently accurate indicators from both the instantaneous and
fossil approach could allow IMF fine-tuning. This approach would
also require an accurate understanding of the dust in star forming
regions, a subject of some controversy in the literature. This topic
will be discussed in a future paper.

\subsection{Dust modelling}
One of the most difficult problems in modelling stellar populations
is how to model the attenuation of the population by dust. For our
previous study \citep{HPJD04} we adopted a simplified model with
only one parameter (the attenuation) while the spectral dependence
of the attenuation was taken to be that of the Large Magellanic
Cloud. Alternate formalisations for the screen are based on the
Small Magellanic Cloud, or estimated for starburst galaxies by
\citet{Calzetti97}. On the other hand, \citet{CF00} have proposed
that a more accurate modelling of the effects of dust attenuation in
galaxies can be achieved by a two-parameter model. In this model one
parameter accounts for dust in the giant molecular clouds
surrounding young stars (of ages$<10$ Myr) and the other parameter
the dust in the diffuse ISM. The combination of the two parameters
allows a more complex extinction curve to be generated.
Unfortunately, since our method does not include the contribution to
the spectrum from emission lines, it is not possible to determine
the extinction from the birth cloud from continuum alone.

%A further method to compensate for dust extinction was given in
%\citet{Mathis06}; a smooth continuum is generated by a 500\AA\
%boxcar, and subtracted from the spectrum, leaving only the high
%frequency signal. Although this approach is certainly effective, it
%removes any chance of recovering information from the shape of the
%continuum, which may provide vital extra information for determining
%the stellar populations of galaxies.

Fig.{\ref{fig:sff}}f and {\ref{fig:sfr_var}}f show the differences
between the LMC, SMC and Calzetti models for the BC03 models at
3\AA\, resolution. The most significant difference occurs clearly at
about $0.1$ Gyr.

\section{Residuals of best fit models}

\begin{figure*}
\includegraphics[width=15cm]{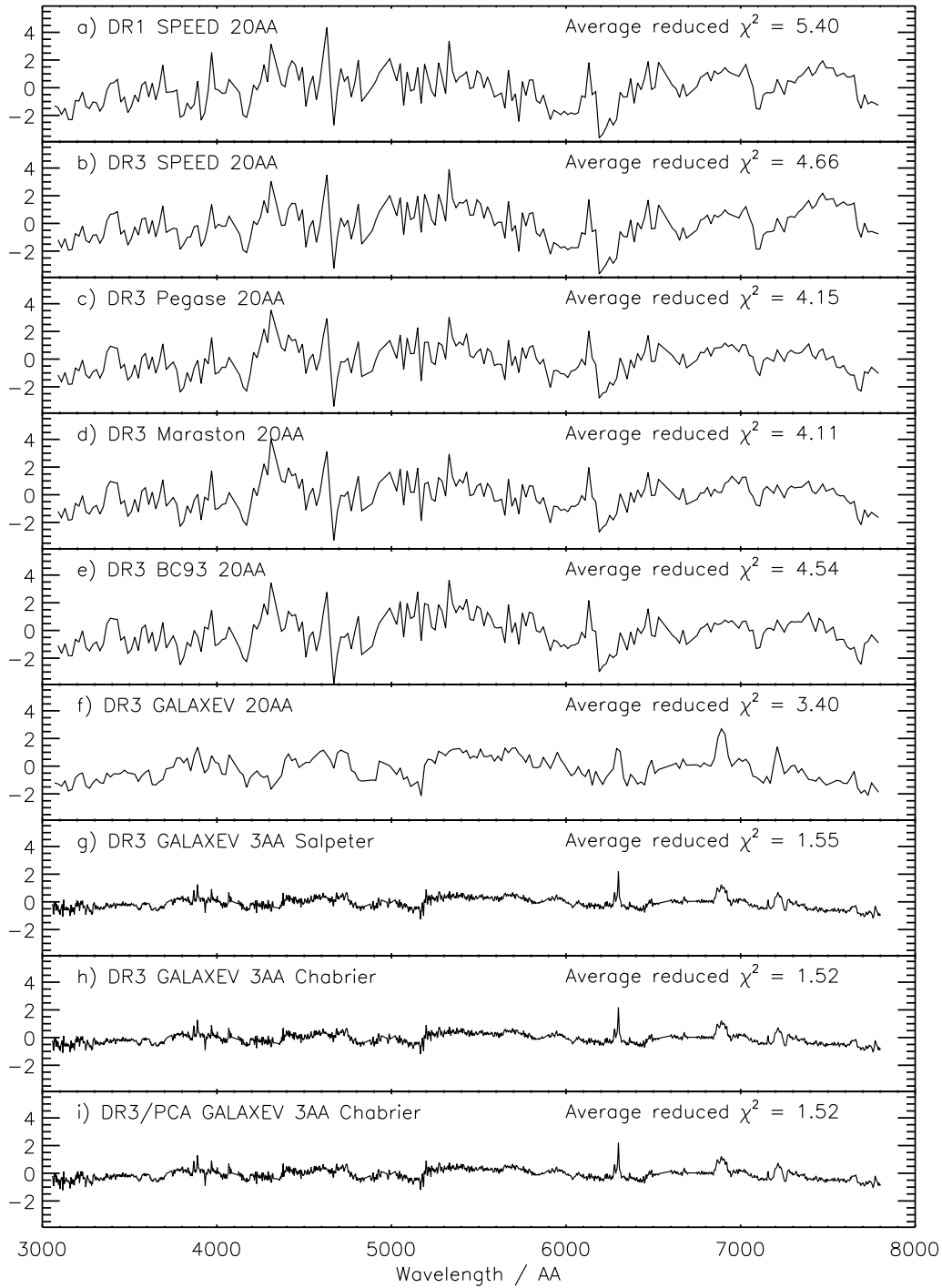}
\caption{Average residuals, following the method used to prepare
figure \ref{fig:dr3_resids} but instead using all the MGS spectra in
the two plates. The individual panels are labelled with the models
and datasets which were used to generate them. From top to bottom,
a) DR1 data, \citet{JMDPP04} SPEED models; b) DR3 data, SPEED
models; c) DR3 data, \citet{Pegase} PEGASE models; d) DR3 data,
\citet{Maraston05} RHB models; e) \citet{BC93} models; f) DR3 data,
\citet{BC03} GALAXEV models rebinned to 20\AA; g) DR3 data, GALAXEV
models at 3\AA\ resolution using a \citet{Salpeter} IMF; h) DR3
data, GALAXEV models at 3\AA\ using a \citet{Chabrier} IMF; i) DR3
data cleaned using the skyline extraction method of \citet{Wild05}.
Residuals are averaged in the rest frame.} \label{fig:res}
\end{figure*}

The average residuals from the best-fit solutions of the different
models is shown in Fig.~\ref{fig:res}. This provides a more detailed
look at how well the models are faring at reproducing the features
in the observed spectra and if some models do better than others. It
can be seen that the 20\AA\ models cover essentially the same range
of spectral features, and it would be impossible to say that one is
better than another. The picture changes when we consider the 3\AA\
models however, they are clearly superior at minimizing the
residuals - even when rebinned to 20\AA\ resolution. Although there
are still several regions where features are not quite correct, it
is on the level of individual lines rather than wide bands of the
spectrum. Comparing panels (a) and (b) allows us to investigate the
effect of the improvement in photometric calibration between DR1 and
DR2-3 on line strengths -- practically none, as it should be.

This average deviation should not be confused with average goodness
of fit however, as inspection of the relevant average $\chi^2$ of
the samples shows a slightly different story. The models have
essentially infinite precision, so there is no penalty associated
with rebinning. The converse is true for the spectra, as binning
pixels while correctly propagating the error will reduce the
standard deviation.

\section{Conclusions}
We have used MOPED to probe the fossil record of star formation
encoded in the spectra of more than $300,000$ SDSS-DR3 galaxies to
determine stellar populations, metallicity evolution and dust
content. We have also investigated the impact of systematics on
recovering physical information from the fossil record.  Our main
conclusions are:
\begin{itemize}
\item The main impacts on systematic variation in the estimation of SFR from
the fossil record are due to the the stellar population model, the
calibration of the observed spectra and the choice of the IMF.

\item We find strong evidence for downsizing, independent of model
choice.

\item The overall star formation history of galaxies recovered from
the fossil record agrees well with instantaneous formation
measurements.

\item The mass build-up recovered from our analysis is
in good agreement with that predicted from both high redshift
studies and the semi-analytic simulations of galaxy formation based
on Millennium Run.

\item We have identified the cause of many of the residuals from the
spectral fits. Stellar population models that provide extra freedom
in terms of alpha enhancement should provide better fits.
\end{itemize}

The fossil record continues to provide a useful tool to unveil the
physical conditions of galaxies in the present and the past. With
improved models that incorporate alpha-enhancement it should be
possible to constrain further models of galaxy formation and
evolution and the initial mass function of galaxies.

\section*{acknowledgments}
We thank the anonymous referee for several suggestions which have
considerably increased the scope and clarity of the paper.

BP and SC thank the Alexander von Humboldt Foundation, the Federal
Ministry of Education and Research, and the Programme for Investment
in the Future (ZIP) of the German Government for funding through a
Sofja Kovalevskaja award. The research of RJ is partially supported
by NSF grants AST-0408698, PIRE-0507768 and NASA grant NNG05GG01G.

BP wishes to thank Paul Hewett for considerable assistance
identifying the source of various residuals in Fig.
\ref{fig:dr3_resids}.

We acknowledge use of the public IDL routines of Craig Mackwardt,
David Fanning and the SDSS idlutils package, and the private
SDSS/Milky Way dust compensation routine of Rita Tojeiro. Much of
the exploratory work that led to the results contained in this paper
benefited from an SQL database created with Gerard Lemson as part of
the GAVO project. We wish to thank Darren Croton for providing the
data used for comparison with the Millenium Simulation in Figure
\ref{fig:sf_build} in electronic form.

Funding for the creation and distribution of the SDSS Archive has
been provided by the Alfred P. Sloan Foundation, the Participating
Institutions, the National Aeronautics and Space Administration, the
National Science Foundation, the U.S.  Department of Energy, the
Japanese Monbukagakusho, and the Max Planck Society. The SDSS Web
site is {\tt http://www.sdss.org/}.

The SDSS is managed by the Astrophysical Research Consortium (ARC)
for the Participating Institutions. The Participating Institutions
are The University of Chicago, Fermilab, the Institute for Advanced
Study, the Japan Participation Group, The Johns Hopkins University,
Los Alamos National Laboratory, the Max-Planck-Institute for
Astronomy (MPIA), the Max-Planck-Institute for Astrophysics (MPA),
New Mexico State University, University of Pittsburgh, Princeton
University, the United States Naval Observatory and the University
of Washington.

%\bibliographystyle{mn2e}
%\bibliography{../../STY/raul}

\end{document}